\documentclass[12pt]{article}
\usepackage{latexsym,epsfig,amssymb, amsmath}
\usepackage{enumerate}

\usepackage{color}

              \newcommand{\rf}[1]{(\ref{#1})}

\def\bfone{\relax{\rm 1\kern-.35em 1}}

\newcommand{\be}{\begin{equation}}
\newcommand{\ee}{\end{equation}}
\newcommand{\ben}{\begin{displaymath}}
\newcommand{\een}{\end{displaymath}}
\newcommand{\bea}{\begin{eqnarray}}
\newcommand{\eea}{\end{eqnarray}}

\newcommand{\bean}{\begin{eqnarray*}}
\newcommand{\eean}{\end{eqnarray*}}

\newcommand{\vp}{\varphi}

\def\Kahler{K\"{a}hler~}

\def\K{K{\"a}hler}

\def\rme{{\rm e}}

\makeatletter \@addtoreset{equation}{section} \makeatother

\begin{document}

\begin{titlepage}

\begin{flushright}
\small ~ \\
\end{flushright}

\bigskip

\begin{center}


{\LARGE \bf  Large Field Inflation and Double $\alpha$-Attractors } \\

\vskip 1.0cm

{\bf Renata Kallosh, Andrei Linde and Diederik Roest} \\

\vskip 0.5cm

{\em Department of Physics and SITP \\ 
Stanford University\\
 Stanford, California
94305 USA\\

{\small {{kallosh@stanford.edu  alinde@stanford.edu}}}} \\

\vskip 0.5cm

{\em Centre for Theoretical Physics \\
 University of Groningen \\
Nijenborgh 4, 9747 AG Groningen, The Netherlands\\
{\small {{d.roest@rug.nl}}}} \\

\end{center}

\vskip 1.5cm

\begin{center} {\bf ABSTRACT}\\[3ex]

\end{center}


We consider a broad class of inflationary models that arise naturally in supergravity. They are defined in terms of a parameter $\alpha$ that determines the curvature and cutoff of these models. As a function of this parameter, we exhibit that the inflationary predictions generically interpolate between two attractor points. At small cutoff $\alpha$, the resulting inflationary model is of plateau-type with $n_s  = 1-2/N$ and $r = 12\alpha/N^{2}$. For $\alpha = 1$, these predictions coincide with predictions of the Starobinsky model and Higgs inflation. In contrast, for large cutoff $\alpha$, the theory asymptotes to quadratic inflation, with $n_s  = 1-2/N$, $r = {8 / N}$.  Both universal predictions can be attributed to a stretching of the moduli space. For intermediate values of $\alpha$, the predictions interpolate between these two critical points, thus covering the sweet spots of both Planck and BICEP2.


\vfill

\end{titlepage}

\vspace{2cm}

\section{Introduction}

Two opposite lines of thought have dominated the development of science at various times, both leading to success when properly applied: 

 1) Use simplest ingredients, assume maximal symmetries, hope for the best, and do not make the theory more complicated than necessary.
 
 2) Use a maximally general approach, do not assume that Nature cooperates until proven otherwise, write all possible terms in the theory and then cut some of them out only if there is a good reason to do so.

The first versions of inflationary theory have been proposed soon after the triumph of the Standard Model and the principle of renormalizability. This principle implied, in particular, that the potential of a scalar field should have the general form including terms $\phi^{n}$ only up to $n = 4$. This seemed to be a reasonable starting point to apply the first of the two approaches mentioned above. It was soon recognized that this approach can be fruitful  if the inflaton field $\phi$ interacts with other fields with small coupling constants $g$. Once one makes this assumption, the theory remains valid until effective masses of the fields become super-Planckian ($g\phi \sim 1$), or the value of the potential energy becomes super-Planckian ($V(\phi) \sim 1$). The simplest models satisfying all desirable criteria could be borrowed from the Standard Model. One could take $V \sim \tfrac12 {m^{2}}\phi^{2}$ or $V \sim \tfrac14 {\lambda}\phi^{4}$, or some combination of these two terms with a (nearly) vanishing vacuum energy, like in the Higgs potential \cite{Linde:1983gd}. 

Yet another idea is to use some kind of global symmetry protecting certain directions of the potential. Several possibilities of this type are discussed in the literature: A shift symmetry $\phi \to \phi+c$ that is broken only by small terms such as ${m^{2}\over 2}\phi^{2}$  \cite{Linde:1983gd}, natural inflation in the axion direction where the flatness of the potential is broken only by some small non-perturbative effects \cite{Freese:1990rb,Kallosh:2014vja}, and axion monodromy potentials in string theory \cite{Silverstein:2008sg}.

These models, as well as many other models of a similar type, often belong to the classes of large field inflation and give predictions compatible with the results of the BICEP2, with $r \sim 0.1$ \cite{Ade:2014xna}, although this is not a general rule.

On the other hand, one can follow the second line of thought and argue that the general inflationary potential can be represented as a series 
\be\label{1}
V(\phi) = \sum_{n= 0}^{\infty} c_{n} \left({\phi\over \Lambda}\right)^{n}  ,
\ee
where $\Lambda$ is some constant which is often identified with the UV cutoff, and $c_{n} = O(1)$ unless there are some reasons why these coefficients are abnormally large or small. The commonly used argument is that $\Lambda$ should be smaller than $M_{p} =1$, so this series diverge at $\phi > 1$, and therefore the behavior of the potential at large $\phi$ is uncontrollable. 
Many authors use it as an argument against large field inflation and in favor of small-field inflation models, with $\phi \ll 1$, and $r\ll 0.1$.

One may argue that the argument given above is a bit naive because the scalar field does not have its own invariant meaning. It enters all physical expressions only due to its contribution to masses of particles with which it interacts. Therefore instead of the naive criterion $\phi < \Lambda \sim 1$ one should have a more sophisticated criterion $g\phi < 1$.  However, in supergravity and string theory the value of the scalar field may have its own geometric interpretation. For example, the $F$-term contribution to the inflaton potential contains the term $e^{K}$. For the simplest \K\ potential $K = \Phi\bar\Phi$ this term is $e^{\Phi\bar\Phi}$, or, restoring the Planck mass,  $e^{\Phi\bar\Phi/M_{p}^{2}}$. Clearly, this means that the value of the scalar field does have meaning in this theory, and it can be measured, just as naively expected, in units of $M_{p} = 1$. This brings us back to the expansion \rf{1}.

But observational data suggest that something is lacking with this argument: BICEP2 as well as Planck to some extent are pointing towards large field inflation. For instance, the simplest theories that provide the best match to the Planck data \cite{Ade:2013uln} are the Starobinsky model \cite{Starobinsky:1980te}  and the Higgs inflation model \cite{Salopek:1988qh}, both of which are large field models, with $\phi \gg 1 $ during the last 60 e-foldings of inflation.

One could consider the existence of these theories just a minor glitch, an exception from the general rule used by many theorists and based on the expansion \rf{1}. However, there is an additional miracle associated with these two classes of models: In the leading approximation in $1/N$, where $N$ is the number of e-foldings, these models give identical predictions for $n_{s}$ and $r$. A subsequent investigation revealed further surprises. Several large classes of theories have been found, all of which have the same observational  predictions in the leading order in $1/N$ \cite{Kallosh:2013hoa,Kallosh:2013tua,Kallosh:2013yoa}. We called these theories ``cosmological attractors.''  Most of these models belong to the class of large field models, disfavored by the arguments based on \rf{1}, and yet all of them predicted the same values of $n_{s}$ and the same (or almost the same) values of $r$ as the models \cite{Starobinsky:1980te,Salopek:1988qh}, in wonderful agreement with the Planck 2013 data. These properties are associated with special features of these theories which could not be envisaged by investigation of random potentials represented by \rf{1}. 

In this paper we highlight a second feature that makes these models even more interesting and challenging: in a different limit,  the predictions of a certain subclass of these models presented in \cite{Kallosh:2013tua, Kallosh:2013yoa} coincide with the predictions of the various versions of chaotic inflation based on monomial potentials, in good agreement with the BICEP2 data.

A natural framework for the investigation of these theories  is the formulation where they have explicit conformal (or superconformal) invariance, which later becomes broken by a specific choice of a gauge. In this formulation, the theories have a cutoff, corresponding to the boundary of the moduli space in terms of the original field variables. This cutoff is also manifest in the Jordan frame, see Section \ref{confattr}. But upon the transformation to the Einstein frame, the position of the cutoff runs away to infinity, when formulated in terms of a canonically normalized inflaton field. This infinite stretching of the field in the vicinity of the boundary of the moduli space (i.e.~near the cutoff) is the main reason for the universality of the observational predictions in a broad class of models of that type \cite{Kallosh:2013hoa, Kallosh:2013tua, Kallosh:2013yoa}. More details of this mechanism can be found in section 5.

To express it in terms similar to those encountered by cosmologists, one may say that the generically non-uniform shape of the potential \rf{1} is analogous to the homogeneity problem. It is solved by exponential expansion of space during  inflation. Similarly, exponential stretching of the field variables upon transition to canonical fields in these models makes the potential \rf{1} exponentially flat, which allows inflation to occur and leads to universal predictions of the theories.

In a certain class of such theories, called $\alpha$-attractors, one can control  the value of the cutoff. The smaller is the cutoff, the faster the predictions of these theories converge to the universal values favored by Planck 2013. But what if we send the value of the cutoff to infinity? In this case, the previous universal behavior disappears, but a new universality becomes manifest: As we will show in the next section, in the limit $\Lambda \to \infty$, the predictions of these theories generically converge towards the predictions of the simplest chaotic inflation model with the potential $\tfrac12 {m^{2}} \phi^{2}$. We will call such theories double attractors. 

The outline of this paper is as follows. In section 2 we discuss the limit $\Lambda \to \infty$ for generic inflationary models. We introduce the role of conformal symmetry and cosmological attractors in section 3. These models are generalized to $\alpha$-attractors in section 4. Section 5 discusses the double attractor nature of this class of models. In section 6 we present more examples. Their appearance in (superconformal) supergravity is the topic of section 7, where two formulations of the theory are given, one is defined on  a disk, the other one on a half-plane. Finally, we offer our conclusions in section 8.

\section{Inflation in the large cutoff limit}\label{largecut}

As we already mentioned, the  Taylor series expansion \rf{1} for the inflationary potential suggests that  the potential is expected to change significantly on scale $\phi \gtrsim \Lambda$ unless there are some reasons to assume that almost all terms in this expansion vanish. 

But what if we interpret recent observational data differently? Prior to the proof of the renormalizability of the electroweak interactions, scientists also indulged in speculations about higher order corrections above the unitarity bound $10^{2}$ GeV. It seemed plausible that something very wrong should happen at high energies, where perturbation theory breaks down. Once the Standard Model of electroweak interactions was constructed and we learned how to make calculations there, suddenly the road was clear for investigations of processes well above the conjectured cutoff at $10^{2}$ GeV, up to the energies approaching the Planck energy.

What if the history repeats itself, and the recent cosmological data can be interpreted as an indication that the Planck energy is not really a limit? Perhaps it is too early to think about it, especially since, as we already mentioned, in many theories  the Planck energy cutoff should not be considered as a cutoff for $\phi$, but a cutoff for masses of the particles interacting with this field. For a moment, we will just ignore this debate and explore a radical possibility: Consider once again the series of the type of \rf{1}, but take a limit $\Lambda \to \infty$. It is a risky exercise, but maybe it will teach us something. 

And indeed, the results are quite intriguing. Large field inflation typically requires knowledge of the potential on scale $\phi \lesssim O(10)$ \cite{Linde:1983gd}. For such fields, all terms higher order in $\phi$ in \rf{1} vanish in the limit $\Lambda \to \infty$; only the lowest terms survive.

\begin{figure}[h!t!]
\vskip -0.2cm 
\centering
\includegraphics[scale=.5]{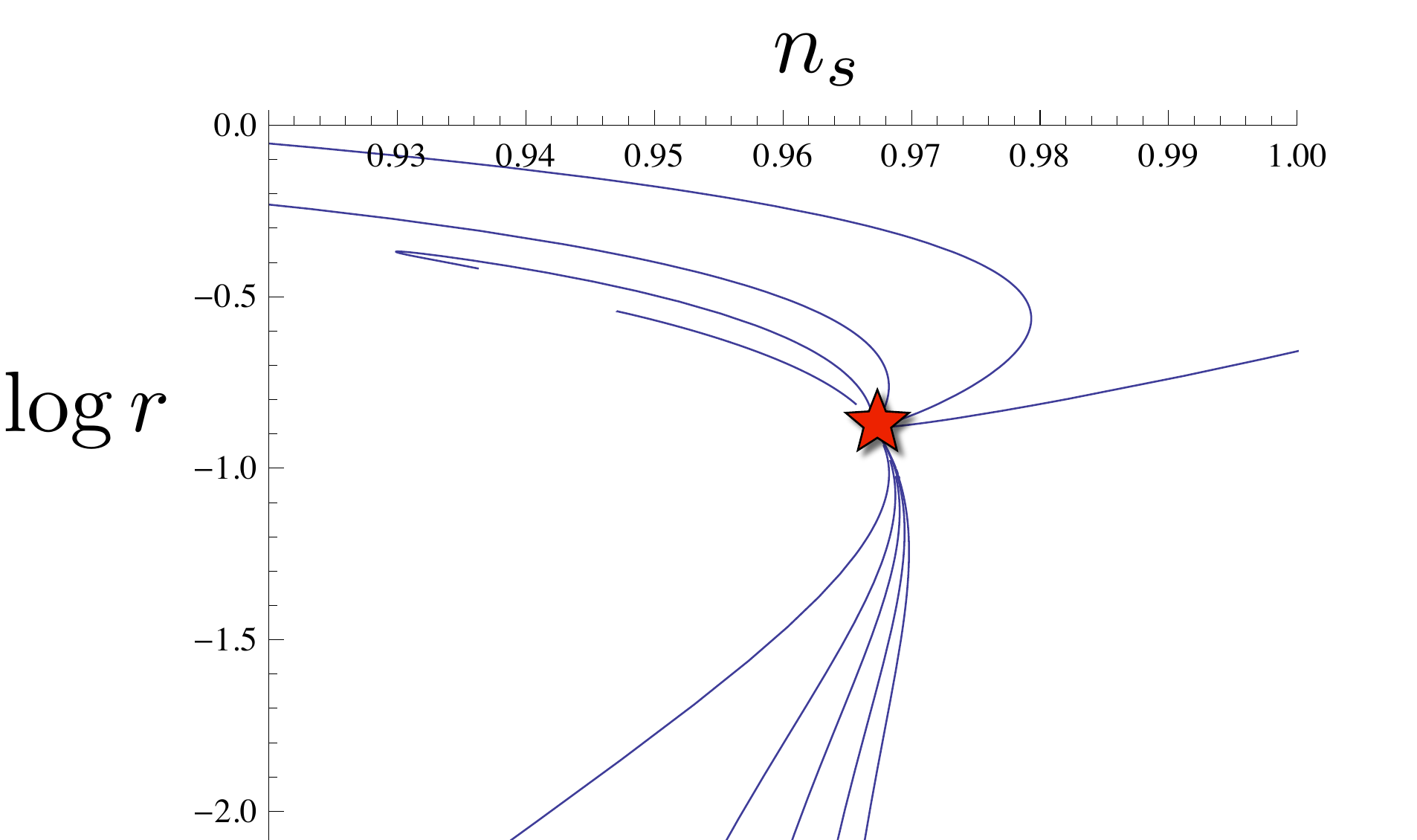} 
\vspace{-.3cm}
\caption{\footnotesize{The cosmological observables $n_s$ and $\log r$ for a representative class of large and small field models with different potentials $V(\phi/\Lambda)$ converge to the predictions of the simplest model with the quadratic potential in the large $\Lambda$ limit. }}
\label{fig:alpha111}
\vspace{-0.3cm}
\end{figure}

The absolute value of the constant term $c_{0}$ in \rf{1} is limited by $10^{{-120}}$ due to anthropic considerations. The linear term exactly
 vanishes at the minimum. Of course, if the coefficient in front of the linear term is smaller than $10^{{-120}}$, we may not be at the minimum as yet (dark energy) \cite{Linde:1986dq}. But this does not change the main conclusion:  Generically, the only term in  \rf{1} that is important for the description of inflationary dynamics at $\phi \lesssim O(10)$ for  $\Lambda \gg 1$ is the quadratic term.
 This means that the predictions of all models \rf{1} should converge to the predictions of the simplest model $\tfrac12 {m^{2}} \phi^{2}$ in the limit $\Lambda \to \infty$. These conclusions can be illustrated by the results of the investigation of the parameters $n_{s}$ and $r$ in a broad class of the models of the type of \rf{1}, see Fig.~\ref{fig:alpha111}.
 
Of course, this conclusion is not absolutely general. For example, the potential may be more complicated so that one cannot even represent it using series \rf{1}, see e.g.  \cite{Silverstein:2008sg} where the potential was quadratic at small $|\phi|$,  and proportional to $|\phi|$ or $|\phi|^{2/3}$ at large $|\phi|$. Our only goal was  to show that if the potential can be represented by the series  \rf{1}, then in the  large $\Lambda$ limit the inflationary potential is quadratic during the last 60 e-folds of inflation.

One could consider this argument just as a mathematical curiosity, but it will play an interesting role in our discussion.
 In the subsequent sections, we will describe a class of large field inflationary models called cosmological $\alpha$-attractors. They exhibit double-attractor behavior.  In a certain  limit, their predictions converge at $n_s  = 1-2/N$ and $r = 0$, which is very close but not exactly the same as in the Starobinsky model. In the opposite limit they also converge at $n_s  = 1-2/N$, but the prediction for $r$ is the same as for the theory $\tfrac12 {m^{2}} \phi^{2}$: $r = {4n /  N}$. As we will see, the predictions span a large area of possibilities on the
 way from one attractor point to another. Moreover, in order to understand the double-attractor nature of these models, we will need to explore both the small and the large cutoff limit.
  
 \section{Conformal symmetry and cosmological attractors}\label{confattr}
 
To explain the structure of the cosmological attractors, we will start with a toy model with the following Lagrangian:
\begin{equation}
\mathcal{L} = \sqrt{-{g}}\left[{1\over 2}\partial_{\mu}\chi \partial^{\mu}\chi  +{ \chi^2\over 12}  R({g})- {1\over 2}\partial_{\mu} \phi\partial^{\mu} \phi   -{\phi^2\over 12}  R({g}) -{\lambda\over 36} (\phi^{2}-\chi^{2})^{2}\right]\,.
\label{toy}
\end{equation}
This
theory is locally conformal invariant under the following
transformations: 
\be \tilde g_{\mu\nu} = \rme^{-2\sigma(x)} g_{\mu\nu}\,
,\qquad \tilde \chi =  \rme^{\sigma(x)} \chi\, ,\qquad \tilde \phi =  \rme^{\sigma(x)}
\phi\ . \label{conf}\ee 
In addition, it has a global $SO(1,1)$ symmetry with respect to a boost between these two fields, preserving the value of $\chi^2-\phi^2$, which resembles the Lorentz symmetry of the theory of special relativity. Note that this theory describes gravity rather than antigravity only for $\chi^2-\phi^2 > 0$. In other words, $\chi$ represents the cutoff for possible values of the field $\phi$.

At the first glance, the physical interpretation of this theory may seem rather obscure, especially because the kinetic term of the field $\chi$ has the wrong sign. This construction could seem `ad hoc', but it is pretty standard; it is commonly used to achieve a mathematically elegant formulation of the standard supergravity.  The field $\chi$, called the conformal compensator or conformon, does not have any physical degrees of freedom associated to it. One can remove it from the theory in several different ways. For example, we may use the gauge $\chi^2-\phi^2=6$ and solve this constraint in terms of the  canonically normalized field $\varphi$: 
$
\chi=\sqrt 6 \cosh  {( \varphi / \sqrt 6)}$, $ \phi= \sqrt 6 \sinh {(\varphi / \sqrt 6)} $.
Our action \rf{toy} becomes
\begin{equation}\label{chaotmodel1}
L = \sqrt{-g} \left[  \frac{1}{2}R - \frac{1}{2}\partial_\mu \varphi \partial^{\mu} \varphi -   \lambda \right].
\end{equation}
Thus our original theory is equivalent to a theory of gravity, a free massless canonically normalized field $\varphi$, and a cosmological constant $\lambda$  \cite{Kallosh:2013hoa}.

As we see, the somewhat unusual term $\sim (\phi^{2}-\chi^{2})^{2}$ in \rf{toy} is essentially the placeholder for what eventually becomes a cosmological constant. The theories to be studied below are based on the idea that one can develop an interesting class of inflationary models by modifying this placeholder, i.e. by locally deforming the would-be cosmological constant. 

Consider a  class of  models
\begin{equation}
\mathcal{L} = \sqrt{-{g}}\left[{1\over 2}\partial_{\mu}\chi \partial^{\mu}\chi  +{ \chi^2\over 12}  R({g})- {1\over 2}\partial_{\mu} \phi\partial^{\mu} \phi   -{\phi^2\over 12}  R({g}) -{1\over 36} f^{2}\left({\phi/\chi}\right)(\phi^{2}-\chi^{2})^{2}\right]\,.
\label{chaotic}
\end{equation}
where $f$ is an arbitrary function of the ratio ${\phi / \chi}$. This theory is invariant under transformations (\ref{conf}), just as the toy model (\ref{toy}). When $f^{2}\left({\phi/\chi}\right)$ is constant, the theory has an additional $SO(1,1)$ symmetry, as we have seen in the example studied above. Introducing a function $f^{2}\left({\phi/\chi}\right)$ is the only possibility to keep the local conformal symmetry (\ref{conf}) and to deform the $SO(1,1)$ symmetry.  The variable $z = {\phi/\chi}$ is the proper variable to describe the shape of the function $f^{2}\left({\phi/\chi}\right)$ in a conformally invariant way.

Using the gauge  $\chi^2-\phi^2=6$, one immediately transforms the theory to the following equivalent form:
\begin{equation}\label{chaotmodel}
L = \sqrt{-g} \left[  \frac{1}{2}R - \frac{1}{2}\partial_\mu \varphi \partial^{\mu} \varphi -   f^{2}(\tanh{\varphi\over \sqrt 6}) \right].
\end{equation}
Note that asymptotically $\tanh (\varphi / \sqrt{6}) \rightarrow \pm 1$ and therefore $f^{2}(\tanh({\varphi / \sqrt 6}))\rightarrow \rm const$; the system  in the large $\varphi$ limit evolves asymptotically  towards its critical point  where the $SO(1,1)$ symmetry is restored.

It is instructive to  consider an alternative derivation of the same result, using the gauge $\chi(x) = \sqrt{6}$ instead of  the gauge $\chi^2-\phi^2=6$. The full
Lagrangian in the Jordan frame becomes
\begin{equation}
\mathcal{L}_{\rm total }= \sqrt{-{g_{J}}}\,\left[{  R({g_{J}})\over 2}\left(1-{ \phi^2\over 6}\right)-  {1\over 2}\partial_{\mu} \phi \partial^{\mu} \phi    -f^{2}\left({\phi\over \sqrt 6}\right) \left({\phi^2\over 6}-1 \right)^{2}\right]\,.
\label{toy2}
\end{equation}
Now one can represent the same theory in the Einstein frame, by changing the metric $g_J$ to a conformally related metric $g_{E}^{\mu\nu} = (1- \phi^{2}/6)^{{-1}}  g_{J}^{\mu\nu}$.
 \begin{align}\label{J1}
  \mathcal{L}= \sqrt{-g} \left[ {1\over 2} R - \frac{(\partial \phi)^2}{ (1- \phi^2 / 6)^2} - f^2( {\phi \over \sqrt{6}})  \right] \, \,.
 \end{align}
Finally, one may express it in terms of a canonically normalized  field $\varphi$ related to the field $\phi$ as follows:
\be\label{field1}
 \frac{d\varphi}{d\phi} = {1\over 1- {\phi^2/ 6}} \, \quad  \Rightarrow \quad
  {\phi \over \sqrt{6}} = \tanh {\varphi \over \sqrt{6}} \,.
\ee
This relates the two Einstein frame formulations \eqref{chaotmodel} and \eqref{J1} to each other.

One can clearly see the presence of the UV cutoff $\Lambda = \sqrt 6$ (i.e. $\Lambda =\sqrt 6 M_{p}$) in \rf{toy2}, \rf{J1}. If the field $\phi$ would become greater than $\sqrt 6$, one would have antigravity instead of gravity. It is natural to expect that something should prevent it from happening  \cite{Kallosh:2013oma,Carrasco:2013hua}. Indeed, the solution  of \rf{field1} shows that while the field $\phi$ approaches the boundary at $\phi = \sqrt 6$, the canonically normalized field $\varphi$ becomes infinitely large. This effect is at the heart of the universality of cosmological predictions in this class of models.

Note that for $|\phi| \ll \sqrt 6$ one has $\phi \approx \varphi$, and the effects of the cutoff can be ignored. On the other hand, once the field $\phi$ approaches the cutoff, the physical distance from the boundary of the moduli space measured by the canonically normalized field $\varphi$ becomes indefinitely large. Unless the function $f(\phi)$ has some very peculiar (singular) behavior near the boundary of the moduli space,  asymptotic behavior of $V(\varphi)$ at large $\varphi$ is universal, which results in the universality of the observational predictions of these models. 
The potentials of the Starobinsky model \cite{Starobinsky:1980te}  and of the the Higgs inflation model \cite{Salopek:1988qh} can be consistently incorporated in this class of models, and the inflaton potentials in these models have the 
functional form which can be cast in the universal form \eqref{chaotmodel}, and they have the same observational predictions:
\be\label{confattra}
n_{s} = 1-\frac2N \,, \qquad  r={12\over N^2} \, .
\ee
For a detailed discussion of related issues and for the description of incorporation of these models to superconformal theory and supergravity see  \cite{Kallosh:2013hoa,Kallosh:2013tua,Kallosh:2013yoa,Kallosh:2013oma,Linde:2014nna} (and \cite{Cai:2014bda} for related work). 

Thus we have a broad class of consistent large field inflation models, which have identical model-independent observational predictions. This universality is closely related to the existence of the UV cutoff in the original conformal formulation of the theory.  Note that we did not need to invent a cutoff, or speculate about its existence as in \rf{1}: It is a part of the theory, and it is directly proportional to the Planck mass. The value of the cutoff becomes infinitely large in terms of the canonically normalized inflaton field in the Einstein frame, but the consequences of its original existence are reflected in the universality of the observational predictions of this class of theories.

In the next section, we will describe $\alpha$-attractors, the theories where the position of the cutoff in the  (super)conformal formulation of the theory can be controlled by some parameter $\alpha$. For each $\alpha \lesssim O(1)$, we will have universal observational predictions, but they will depend on the value of $\alpha$.
 
\section{Cosmological $\alpha$-attractors}

As we will see, the cosmological $\alpha$-attractors can be introduced in several inequivalent ways, but at the end of the day, they lead to the Einstein frame Lagrangian for the inflaton field which can be written in the form similar to  \rf{J1}:
 \begin{align}
  \mathcal{L}= \sqrt{-g} \left[ {1\over 2} R - \frac{\alpha (\partial \tilde\phi)^2}{ (1- \tilde\phi^2 / 6)^2} - f^2(\tilde\phi / \sqrt{6})  \right],
\label{a} \end{align}
 or, equivalently, after rescaling of the field $\phi$, as
  \begin{align}\label{newJordan}
  \mathcal{L}= \sqrt{-g} \left[ {1\over 2} R - \frac{(\partial \phi)^2}{ (1- \phi^2 / (6\alpha ))^2} - f^2(\phi / \sqrt{6\alpha})  \right].
 \end{align}
Upon canonically normalising the kinetic terms for the scalar field via the redefinition
 \begin{align}
 {\phi \over \sqrt{6 \alpha}} = \tanh {\varphi \over \sqrt{6 \alpha}} \,,
 \end{align}
this leads to the Einstein frame $\alpha$-attractor theory:
 \begin{align}
  \mathcal{L}_{\alpha} = \sqrt{-g} \left[ {1\over 2} R - {1\over 2}  (\partial \varphi)^2 -   f^2\big(\tanh {\varphi\over\sqrt{6\alpha}}\big)
 \right].
 \label{action1}  
 \end{align} 
Note that the position of the cutoff in  \rf{newJordan} changed as compared to \rf{J1}: $\Lambda = \sqrt{6\alpha}$. Thus we can control the position of the cutoff by changing $\alpha$, which leads to a broad set of possibilities. 
    
In the subsequent sections, we will discuss the derivation and interpretation of this class of models in the context of superconformal theory and supergravity. Here we will concentrate on their observational consequences:
 \begin{itemize} 
 \item For $\alpha = 1$, the predictions of this class of models coincide at leading order in $1/N$ with the predictions of the Starobinsky model, the Higgs inflation, and the large class of conformal attractor models discussed in the previous section, see equation \rf{confattra}.  
 \item For $\alpha \lesssim 1$, $N \gg 1$ and for generic functions $f(x)$, this class of models leads to a universal prediction \cite{Ferrara:2013rsa,Kallosh:2013yoa}  
\be\label{aattr}
n_{s} = 1- \frac2N \,, \qquad r= {12 \alpha \over N^2} \,,
\ee
while subleading corrections will be model-dependent \cite{Roest:2013fha}. One of the features of the universal attractor here is the absence  of dependence on the choice of $f(x)$ in these results. 
 \item
In the limit $\alpha \to 0$ one finds the universal attractor prediction
\be
n_{s} = 1- \frac2N \,, \qquad r=0 \, .
\ee
with identical subleading corrections for all models.
 \end{itemize}

 \begin{figure}[h!]
\centering
\includegraphics[scale=.45]{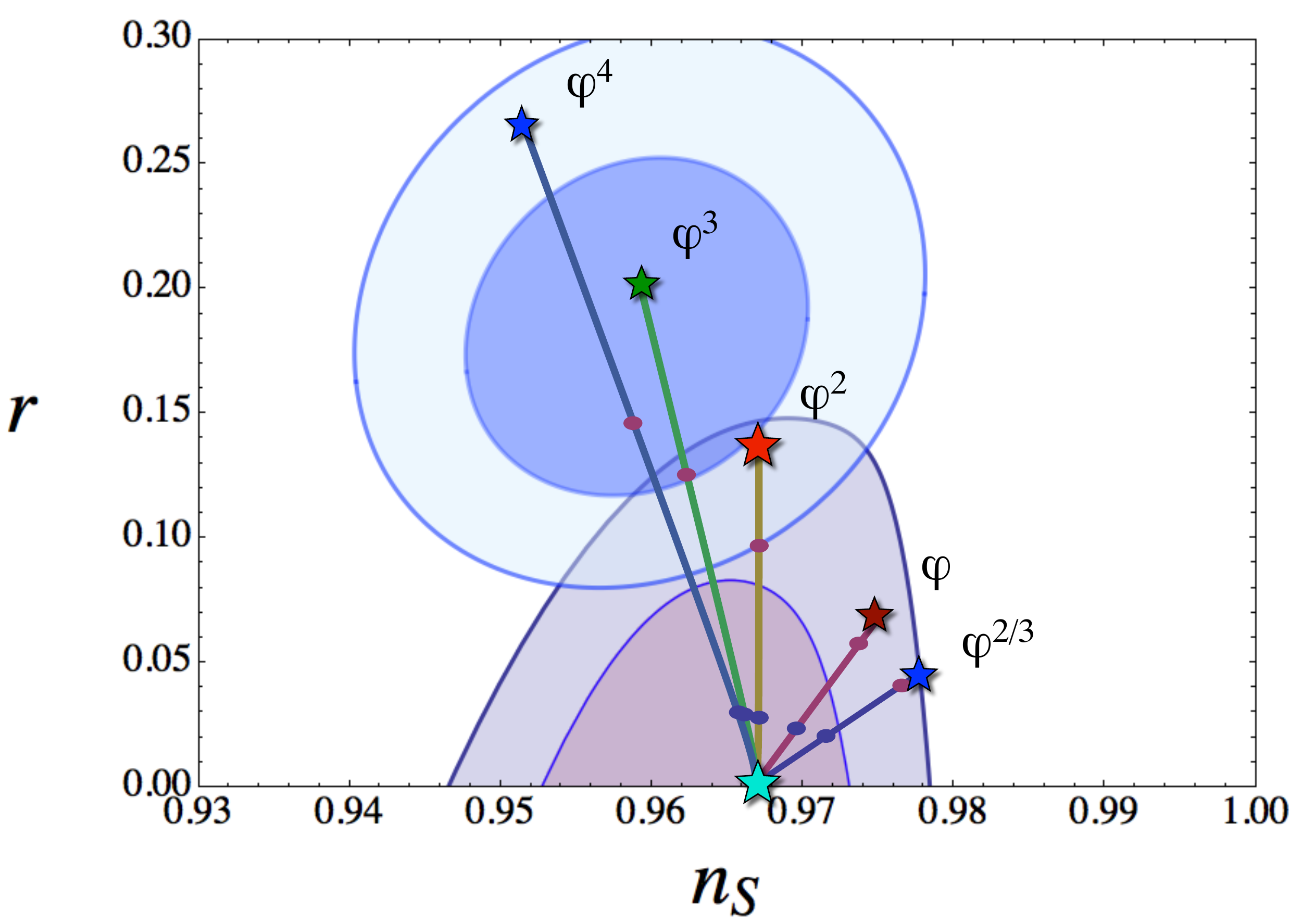} 
\caption{\footnotesize{ The cosmological observables $(n_s,r)$  for different scalar potentials $\tanh^{2n} ({\vp \over \sqrt{6 \alpha}})$ with $2n = (2/3, 1, 2, 3, 4)$  for $N=60$. These continuously interpolate between the predictions of the simplest inflationary models with the monomial potentials $\varphi^{2n}$ for $\alpha \rightarrow \infty$, and the attractor point $n_{s} =1-2/N$, $r = 0$ for $\alpha \to 0$, shown by the bright blue star. The different trajectories  form a fan-like structure for $\alpha \gg n^2$. The set of dark red dots at the upper parts of the interpolating straight lines corresponds to $\alpha = 100$. The set of dark blue dots corresponds to $\alpha = 10$. The lines gradually merge for $\alpha = O(1)$.  The upper blue contours correspond to BICEP2 results, the lower contours correspond to Planck 2013.}}
\label{fig:ClassI}
\vspace{-.3cm}
\end{figure} 

Thus for sufficiently small $\alpha$ we have a universal behavior, independent of the function $f$. This implies that, at a given small value of $\alpha$ there are many models which have the same values of $n_s$ and $r$. For $\alpha=1$ this includes the Starobinsky and Higgs model \cite{Starobinsky:1980te, Salopek:1988qh} but the same universality holds for any value of $\alpha$. Decreasing $\alpha$, one can reach arbitrarily small values of gravity waves, without a significant change in $n_s$.

In the opposite limit of large $\alpha$, we may encounter a full spectrum of possibilities instead. As an example, we have considered monomial functions $f(x) = x^{n}$ \cite{Kallosh:2013yoa}. At small $\alpha$, we have the same universal behavior \rf{aattr}. In the large  $\alpha$ limit one has the predictions coinciding with the predictions of the simplest chaotic inflation with $V \sim \phi^{2n}$, given by
\be\label{largealpha}
n_{s} =1-\frac{2}{N}, \qquad r = {8n\over N}. 
\ee
The trajectories connecting these two limiting regions are almost exactly straight. A full numerical investigation reveals the  fan-like picture with a universal  attractor at small $\alpha$ and $n$-dependent split at large $\alpha$, as presented  at Fig 2. in  \cite{Kallosh:2013yoa}, and reproduced here in Fig.~\ref{fig:ClassI}.

In the following section, we will consider the limit of large-$\alpha$ for  generic functions $f$, and discuss the emergence of a second attractor in this limit.

\section{A second attractor at large $\alpha$}

In  \cite{Kallosh:2013yoa} we studied the limit of large $\alpha$ for the choice $f(x)= x^n$. We found an $n$-dependent split of trajectories of the fan-type, leading to $\phi^n$ chaotic models as shown here in Fig.~2. Now we propose to consider   models \rf{action1} with a general class of functions $f(x)$  defined by the Taylor series similar to \rf{1}:
\be
f(x)_{\rm general} = \sum_n c_n x^n, \qquad n = 1, 2, 3,... \ .
\ee
In general, none of the coefficients $c_n$ is expected to be zero. However, in this investigation we will concentrate on the models which have a minimum of the potential at $x = 0$.  Then $c_{0}$ must be smaller than $10^{-60}$ because of anthropic considerations, see a similar argument in Section \ref{largecut}.

\begin{figure}[h!t!]
\centering
\includegraphics[scale=.45]{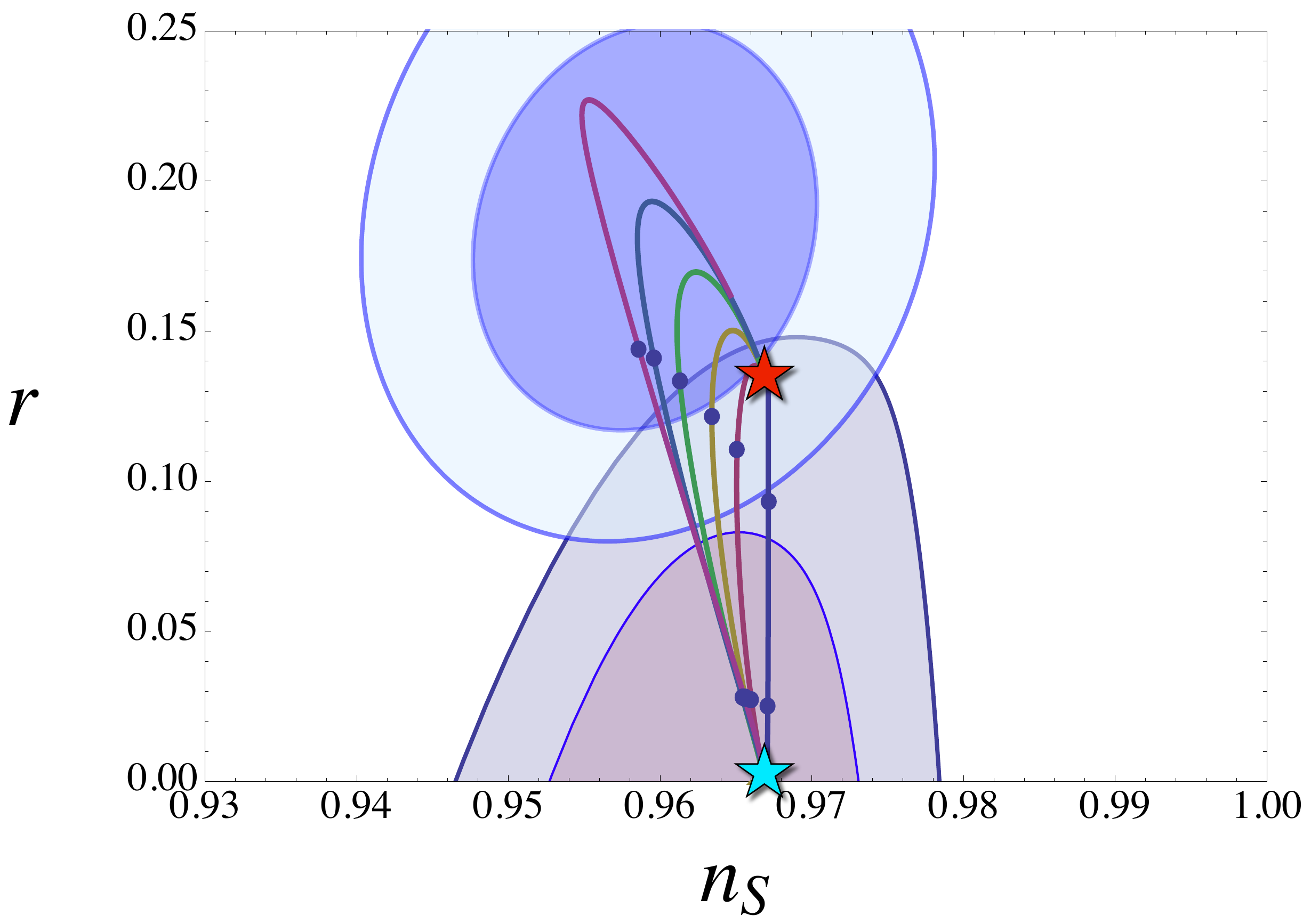} 
\caption{\footnotesize{The double $\alpha$-attractor  in the $(n_s, r)$ plane for different chaotic models with $f(x)= x+c x^2$ with $c=(0, 0.4, 0.8, 1.6, 3.2, 10)$ (in order from right to the left) for 60 e-folds and $M=1$. The dots correspond to $\log \alpha = (1,2)$. The dots which would correspond to $\alpha = 1$ practically merge with each other and are covered by the bright blue star corresponding to $\alpha = 0$. The lower attractor at small $\alpha$ is at small $r$, the upper attractor at large $\alpha$ is at the point where all models lead to a $\phi^2$ model. }}\label{fig:alpha}
\vspace{-.3cm}
\end{figure} 

At large-$\alpha$ one can use the approximation  $\tanh ( {\varphi / \sqrt{6\alpha}})\, \approx  \, {\varphi / \sqrt{6\alpha}}$, which holds when $\vp\ll \sqrt {6\alpha}$.
In this limit we have $f\big(\tanh ( {\varphi / \sqrt{6\alpha}}) \big) = c_{1 }{\varphi / \sqrt{6\alpha}}\ll 1$, and therefore 
\begin{align}
V(\vp) =    f^2\big(\tanh {\varphi\over\sqrt{6\alpha}}\big) = {c_1^{2}\over 6\alpha} \varphi^2  \, .
 \label{actionLim}  \end{align} 
Note that in the purely quadratic chaotic inflation model one has $N = \vp^{2}/4$ \cite{Linde:1983gd}.  Therefore one can self-consistently describe inflation in the quadratic approximation \rf{alphanew2} for  $\alpha \gg 2N/3$, which yields the constraint $\alpha \gg 40$ for $N = 60$. 

An exception to this conclusion is provided by the monomial examples $f= x^n$ with $n \neq 1$ studied in \cite{Kallosh:2013yoa}; these do not have a quadratic expansion \eqref{actionLim} for any value of the stretching parameter $\alpha$, see Fig.~\ref{fig:ClassI}. However, in reality we do expect the potential to be quadratic near its minimum. Therefore, in the large $\alpha$ limit, the last 60 e-folds of inflation always occur near the minimum where the potential is quadratic.

To illustrate the emergence of the attractor regime in the large $\alpha$ limit, we considered functions $f(x) = x+c x^2$ and plotted $n_{s}$ and $r$ for various values of  $c$ and for $\alpha$ interpolating between 1, 10, 100 and  higher values as shown in Fig.~\ref{fig:alpha}. We have studied other examples, like $f(x) = x+c x^3$ and found analogous behavior at large $\alpha$. As expected, we have found a double-attractor regime, with predictions of a broad family of models continuously interpolating between the set of data favored by Planck 2013 and the data favored by BICEP2.

\begin{figure}[h!t!]
\centering
\includegraphics[scale=.5]{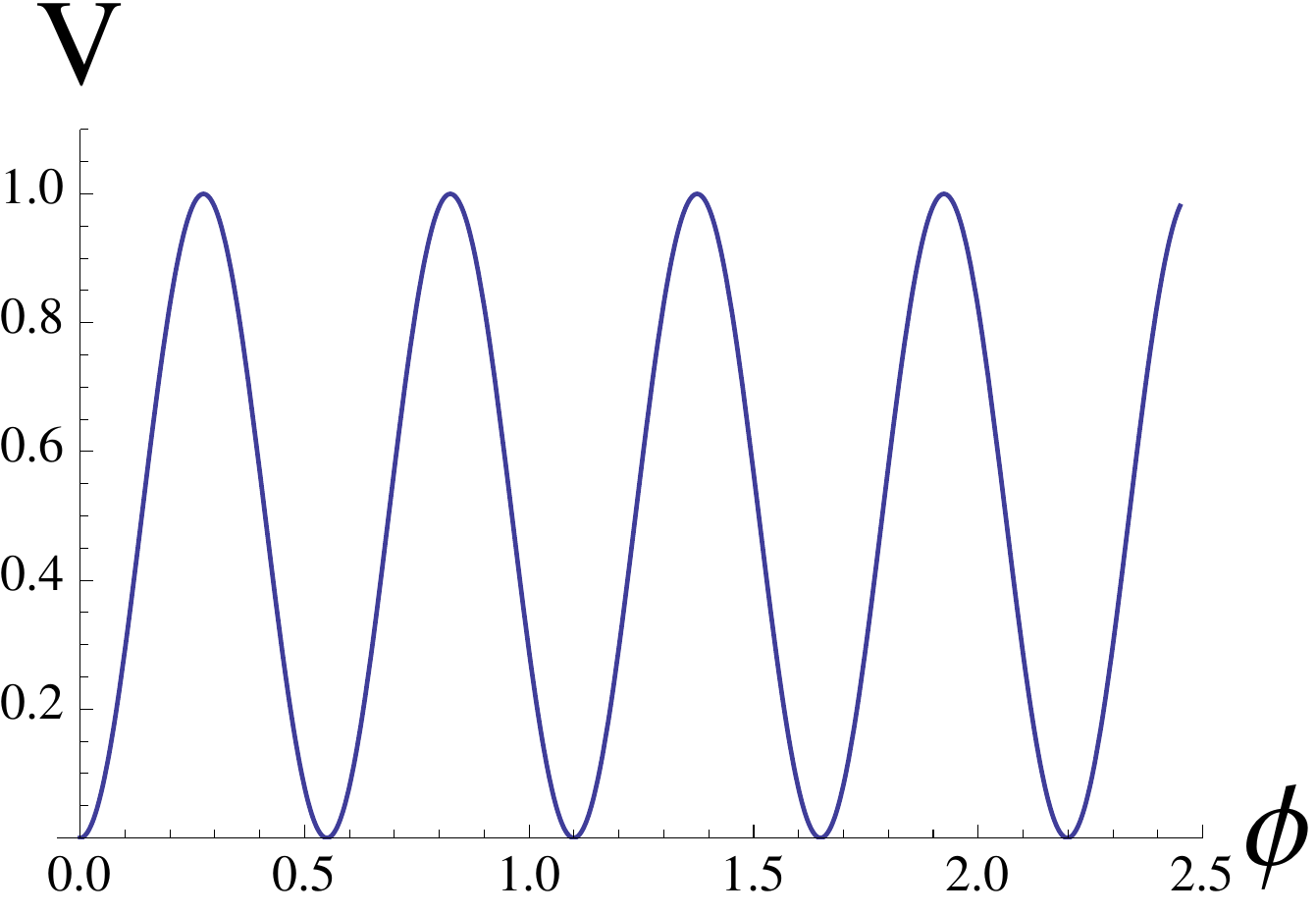} \;\;\;\;\; 
\includegraphics[scale=.5]{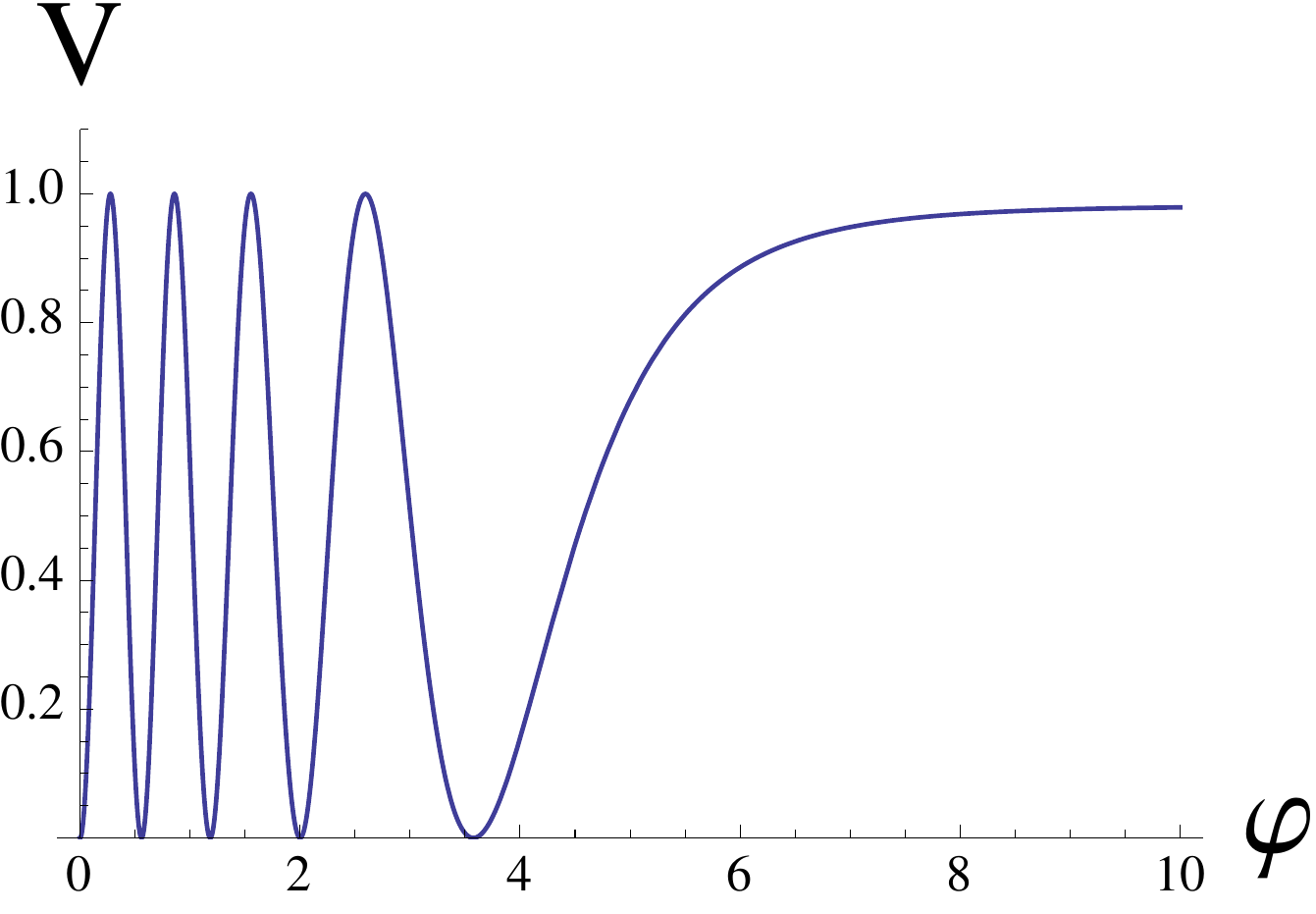}
\includegraphics[scale=.49]{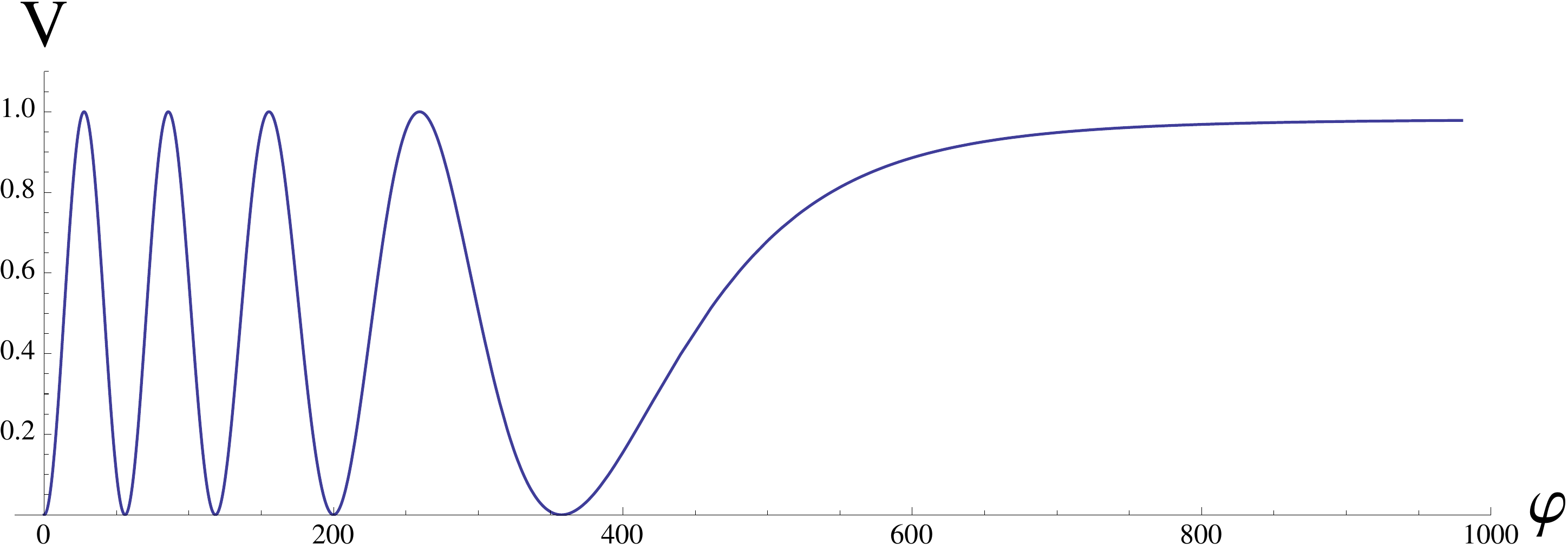}  
\caption{\footnotesize{a) A periodic potential in the theory \rf{newJordan} for $\alpha = 1$;  b) The same potential in terms of the canonically normalized field $\varphi$ for $\alpha = 1$. The emergence of an infinitely long plateau of the potential, with a sufficiently sharp fall-off, is responsible for the attractor behavior of this class of models for $\alpha \lesssim 1$; c) The same potential in terms of the canonically normalized field $\varphi$ for $\alpha = 10^{4}$. The uniform horizontal stretching of the inflationary potential is responsible for the attractor behavior in this class of theories in the limit $\alpha\to \infty$.  Because of the stretching of the potential, it became very flat. Therefore inflation may occur near each of the many minima of the potential.}}  \label{fig:1a}
\vspace{-.3cm}
\end{figure}

Here we would like to provide a simple graphical illustration explaining the origin of the double-attractor regime in this class of models. Fig.~\ref{fig:1a}a shows the potential of the theory \rf{newJordan} with a simple sinusoidal potential of the field $\phi$. The potential is periodic with a period smaller than 1, which means that if the field $\phi$ were canonically normalized, there would be no inflation in this theory. 

However, in terms of a canonically normalized field $\vp$, the potential acquires an infinitely long flat direction, which makes inflation possible, see Fig.~\ref{fig:1a}b, which shows the potential for $\alpha = 1$. The inflationary regime naturally emerges in this theory even if the original potential in the theory \rf{newJordan} is very curved. Importantly, the potential at large $\vp$ has a universal shape which is almost independent on the detailed behavior of the inflaton potential in  \rf{newJordan} near the boundary of the moduli space at $\phi = \sqrt 6$. This is the reason for the universality of the observational predictions in this class of theories.

Finally, Fig.~\ref{fig:1a}c shows the potential for $\alpha = 10^{4}$. The only difference between this potential and the potential in Fig.~\ref{fig:1a}b is that the new potential is stretched horizontally by a factor of 100. As a result, the curvature of the potential decreases, and now inflation may occur in the vicinity of each of the many minima of this potential, which were unsuitable for inflation for $\alpha \lesssim 1$. The inflationary regime at the infinitely long plateau remains possible, but the last 60 e-foldings of inflation occur in a relatively small vicinity of each minimum, at a distance $\Delta \phi \sim 15$ from each minimum. Since in the vicinity of each minimum, the potential is approximately quadratic, the predictions of the inflationary regime near each minimum practically coincide with the predictions of the simplest version of chaotic inflation with a quadratic potential.

\section{Other examples of $\alpha$-attractors}

Since one has full freedom of choice of the functions $f(\phi / \sqrt{6\alpha})$, one can also represent a family of the inflaton potentials belonging to this class using some different function  $\tilde f({\phi / \sqrt{6\alpha}\over 1+\phi / \sqrt{6\alpha}})$. In terms of the canonical variables, this theory looks as follows:
\begin{align}
  \mathcal{L}_{\alpha} = \sqrt{-g} \left[ {1\over 2} R - {1\over 2}  (\partial \varphi)^2 -   {\tilde f}^2\big(1-e^{-\sqrt{3\over 2\alpha}\vp}\big)
 \right].
 \label{action111}  \end{align} 
 For the simplest choice of the function $ {\tilde f}(x) \sim x$, this model has a potential 
  \be\label{alphanew1a}
V= V_{0}\big(1- e^{-\sqrt {2\over 3\alpha} \varphi}\big)^2 \ .
\ee
This is a particular example of the supersymmetric $\alpha$-model discovered in \cite{Ferrara:2013rsa}, which for $\alpha = 1$ coincides with the potential  of the Starobinsky model \cite{Starobinsky:1980te}. We have studied the $\alpha$-dependence of the cosmological predictions of this model in \cite{Kallosh:2013yoa}. For small $\alpha$ the cosmological predictions of this model were studied in \cite{Ellis:2013nxa} and in \cite{Ferrara:2013rsa}.

When $\alpha$ becomes very large, one has
\be\label{alphanew2}
V= V_{0}\big(1- e^{-\sqrt {2\over 3\alpha} \varphi}\big)^2 \approx {m^{2}\over 2}\varphi^{2} \ .
\ee
where $m^{2} = {4V_{0} / (3\alpha)}$. 
This approximation is valid for $\varphi \ll \sqrt {3\alpha / 2}$. As we already mentioned, in the purely quadratic chaotic inflation model one has $N = \vp^{2}/4$. Therefore one can self-consistently describe inflation in the quadratic approximation \rf{alphanew2} for  $\alpha \gg 8N/3$, which yields the constraint $\alpha \gg 160$ for $N = 60$. In the large $N$ limit, in the quadratic approximation one has 
\be
n_{s} =1-\frac{2}{N}, \qquad r = {8\over N}. 
\ee

\begin{figure}[h!t!]
\vskip -0.2cm 
\centering
\includegraphics[scale=.38]{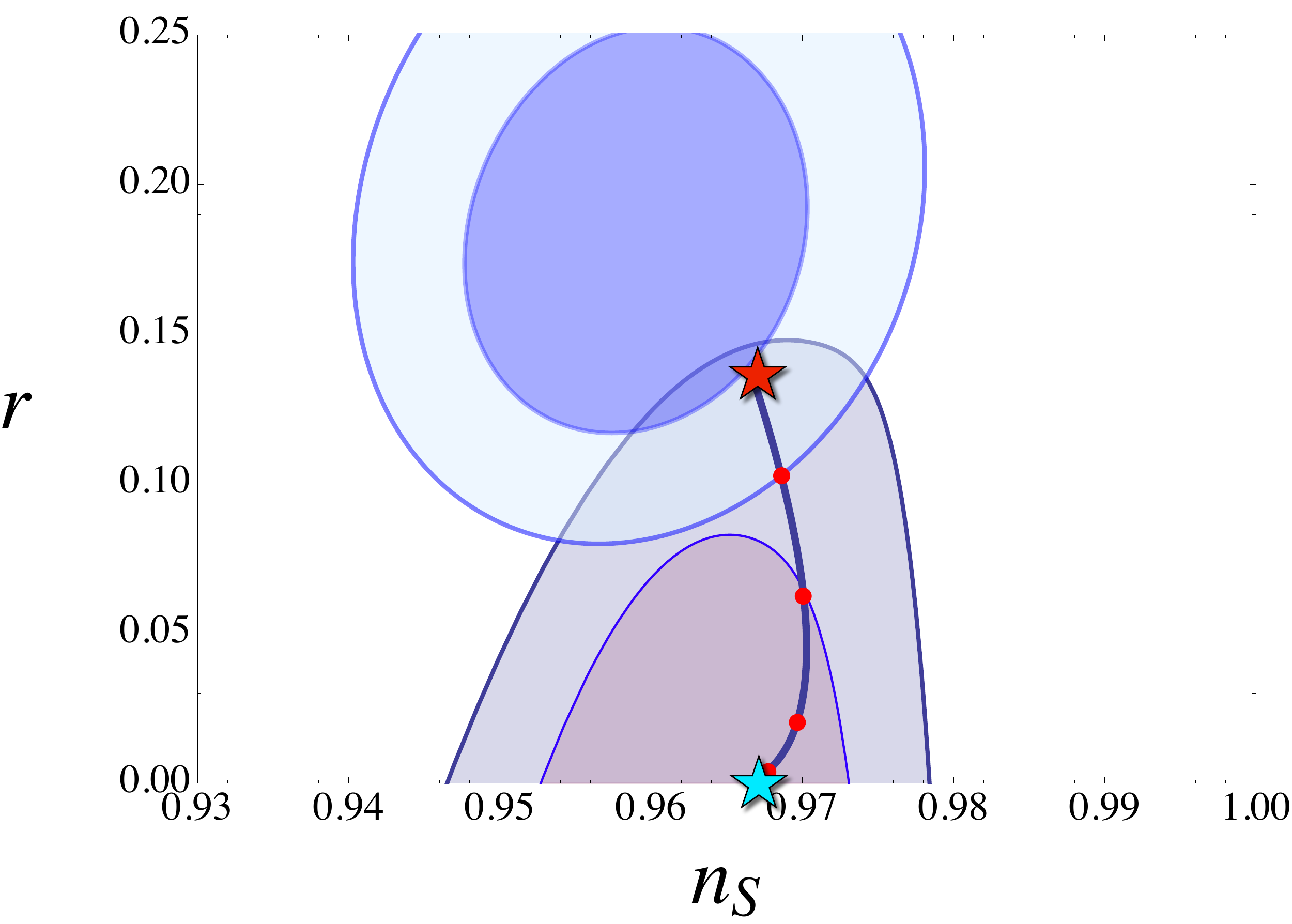} 
\vspace{-.3cm}
\caption{\footnotesize{The cosmological observables $n_s$ and $r$ for the theory with a potential $V_{0} \Bigl(1- e^{-\sqrt {2\over 3\alpha} \varphi}\Bigr)^2$ for $N=60$. As shown by the thick blue line, $n_s$ and $r$ for this model depend on $\alpha$ and continuously interpolate between the prediction of the simplest chaotic inflationary model with $V \sim \varphi^{2}$ for $\alpha \rightarrow \infty$ (red star), the prediction of the Starobinsky model for $\alpha = 1$ (the lowest red dot), and the prediction $n_{s} =1-2/N$, $r = 0$ for $\alpha \to 0$ (blue star). The red dots on the thick blue line correspond to $\log \alpha = \{3, \ldots, 0\}$, from the top down. }}
\label{fig:alpha1}
\vspace{-0.3cm}
\end{figure} 

By continuously decreasing $\alpha$ from $\infty$ to $0$, one can cover the full range of possible values of $r$ from $r = {8/N}$ to $r = 0$. The last part of this trajectory, when $\alpha$ is of order one or smaller, proceeds along the attractor regime discussed above for small $\alpha$. The results of a numerical investigation of the parameters $n_{s}$ and $r$ in this model  are represented by a thick blue line in Fig.~\ref{fig:alpha1}. 

What will happen if one considers the same model \rf{action111}, but with $\tilde f(x) \sim x^{n}$? In the small $\alpha$ limit we will have the same universal predictions \rf{aattr}. But, just as in the case studied in the previous sections, the situation changes dramatically in the large $\alpha$ limit. For $\varphi \ll \sqrt {3\alpha / 2}$ the potential will behave as $\vp^{2n}$, with observational predictions (for  $N \gg n$) given by \eqref{largealpha}. The resulting picture will therefore be similar to the fan-like behavior of Fig.~\ref{fig:ClassI}, but with curved instead of straight lines.

Similarly, choosing a more generic function $\tilde f(x) \sim x + c x^2+...$ will provide an interpolation between the two attractor points similar to Fig.~\ref{fig:alpha}, but again with a different shape of lines in-between the two attractor points. 

In the following sections, we will describe the implementation of the double-attractor models in the context of superconformal theory and supergravity. We will see that the cutoff controlled by $\alpha$ has an invariant geometric interpretation in these theories as a measure of the curvature of the \K\ manifold.

\section{Superconformal $\alpha$-attractors}

The supersymmetric $\alpha$-attractor  models were proposed in \cite{Kallosh:2013yoa}, following the first supersymmetric cosmological model \rf{action111} with $\alpha\neq 1$, which  was found in \cite{Ferrara:2013rsa}, where also the cosmology of these models was studied in the small $\alpha$ limit.  All  models which we discuss below correspond to ${\cal N}=1$ supergravity embeddings of bosonic  inflationary models. It is interesting that  recently a new  ${\cal N}=2$ supergravity embedding of the inflationary models \rf{alphanew1a} was achieved in \cite{Ceresole:2014vpa} for the  choices  $\alpha=1/3, 2/3$ and $1$.

\subsection{The model on a disk}

At the superconformal level we  consider 3 chiral supermultiplets: a conformon $X^0$,  an inflaton $X^1=\Phi$ and a goldstino multiplet $X^2=S$, which has a first component scalar, the sgoldstino. The superconformal models are defined by two arbitrary functions of these superfields.  The first function is a \Kahler potential of the embedding manifold ${\cal N} (X, \bar X)$. It is real and has a Weyl weight 2. The second one represents a superpotential ${\cal W}(X)$. It is holomorphic and has a Weyl weight 3. The Lagrangian in terms of these functions is
 \begin{align}
  \mathcal{L} = \sqrt{-g} \left[ - \tfrac16 \mathcal{N}(X, \bar X) R - G_{I \bar J} \mathcal{D}^\mu X^I \mathcal{D}_\mu {\bar X}^{\bar J} - G^{I \bar J} \mathcal{W}_I \mathcal{\bar W}_{\bar J}
\right] \,,
 \end{align}
with $I, \bar I = \{ 0,1,2 \}$. The superconformal \Kahler potential for the $\alpha$-attractor models is  given by \cite{Kallosh:2013yoa}
 \begin{align}
  \mathcal{N}(X, \bar X)  = - |X^0|^2 \left[ 1 - \frac{|X^1|^2 + |S|^2}{|X^0|^2} \right]^\alpha \,.
 \end{align}
Note that the \Kahler potential only preserves the manifest $SU(1,1)$ symmetry between $X^0$ and $X^1$  for the special value $\alpha = 1$. The superconformal superpotential reads
 \begin{align}
  \mathcal{W} = S (X^0)^2 f(X^1 / X^0) \left[ 1 - \frac{(X^1)^2}{(X^0)^2} \right]^{(3 \alpha -1)/2} \,.
\label{calW} \end{align}
 The superpotential with a constant $f$ and $\alpha =1$ preserves the $SO(1,1)$ symmetry, the subgroup of $SU(1,1)$. However, when either $f$ is not constant, or $\alpha \neq 1$,  the $SO(1,1)$ symmetry is deformed.
In order to extract a Poincar{\'e} supergravity we gauge fix the conformal symmetry by setting $X^0 = {\bar X}^0 = \sqrt{3}$. This leads to a 
 supergravity version of the model described in \cite{Kallosh:2013yoa} is given by the 
 \Kahler and superpotential 
  \begin{align}
  K & = - 3 \alpha \log \left[1 - |Z|^2 - \frac{S \bar S}{3}  + \frac{g (S \bar S)^2}{3 (1 - |Z|^2)} \right] \,, \qquad
  W = S f(Z) (1 - Z^2)^{(3 \alpha -1)/2} \,.
\label{KW} \end{align}
Here $g$ is the constant coefficient of a stabilization term that is required for small $\alpha$ \cite{Kallosh:2013yoa}, but will play little role in what follows.

At $S=0$ the complex variable $Z= X^1/X^0$ is restricted to a disk
\be
 |Z|^2 < 1
\ee
which is a boundary of the moduli space. The action at $S=0$ is
 \begin{align}
  \mathcal{L} = \sqrt{-g} \left[ {1\over 2} R - 3  \alpha \frac{ \partial Z \partial \bar Z}{  (1- |Z|^2 )^2} - f^2(Z)  \right] \, \,.
 \end{align}
 The \K\, geometry in this theory {\it for any value of $\alpha$ has an $SU(1,1)$ symmetry} associated with the  symmetry of the kinetic term. We will describe it in detail in the next subsection where it becomes an $SL(2,  \mathbb{R})$ symmetry.

 For any real functions $f$, the   model above allows for a truncation to a one-field model via $S = Z - \bar Z = 0$. The  effective Lagrangian at $S = Z - \bar Z = 0$ is
\begin{align}
  \mathcal{L} = \sqrt{-g} \left[ {1\over 2} R - 3  \alpha \frac{ (\partial Z)^2}{  (1- Z^2 )^2} - f^2(Z)  \right] \, \,.
 \end{align} 
 Thus the action is greatly simplified for real $Z$. This form of the stabilized supergravity action coincides with the bosonic action \rf{a} and explains the supersymmetric origin of the parameter $\alpha$ in the bosonic model  \rf{a} .

One can see the  role of the parameter $\alpha$ by performing the  change of 
$ Z  = {\phi / \sqrt  {6 \alpha}}$ to
 \begin{align}
  \mathcal{L} = \sqrt{-g} \left[ {1\over 2} R - {1\over 2} \frac{(\partial \phi)^2}{  (1- {\phi^2\over  6\alpha} )^2} - V\Big ({\phi\over \sqrt {6\alpha}}\Big )  \right] \, \,.
 \end{align}
There is  a simple relation between the geometric field $Z= {\phi / \sqrt  {6 \alpha}} $ at $Z=\bar Z$ and a canonical one $\varphi$: it is the rapidity-like relation 
\be
Z =  {\phi \over \sqrt  {6 \alpha}}  = \tanh {\varphi \over \sqrt {6 \alpha}} \, 
\label{rapidity}
\ee
 where the geometric field $Z$ has to be inside the disk $ |Z|^2 < 1$, implying that 
$\phi^2 < 6 \alpha$, whereas the canonical field $\varphi$ is unrestricted.

\subsection{The model of the half-plane} 
Here we will describe the same model as given in \cite{Cecotti:2014ipa}, after the map of the disk to the half-plane.  The model on the disk in in \cite{Kallosh:2013yoa} is given in eq. \rf{KW}  and it becomes more elegant when transformed to the half-plane by means of 
\be
Z= {T-1\over T+1}  
\ee
and for real fields $ {\rm Re}\,  Z\rightarrow 1 \Rightarrow  {\rm Re}\,  T\rightarrow \infty $  and ${\rm Re}\,  Z\rightarrow-1 \Rightarrow {\rm Re}\,  T\rightarrow 0$. In these variables our 
 supergravity model \rf{KW} becomes,  up to a \Kahler transformation and overall factor for the superpotential,
\be
K= -3\, \alpha \log \left(T + \bar T   -  C \bar C  
 + 3 g \frac{( C \bar C)^2}{T + \bar T}\right)\, , \qquad 
 W=  C F(T) \ .
\label{sugra}\ee
Here the original $S$ variable is related to $C$ as follows: $S= \sqrt 6 {C\over T+1}$.
The relation to the function $f(Z)$ in \cite{Kallosh:2013yoa} is explained in details in \cite{Cecotti:2014ipa}. It is given by the following formula 
\be
F(T)\equiv  T^{(3 \alpha -1)/2} f\Big ( {T-1\over T+1}\Big )= T^{(3 \alpha -1)/2} \tilde f(T) \ .
\label{Cecotti}\ee 
The bosonic part of the supergravity model  at the minimum at $C=0$  is given by the following expression 
\be
e^{-1} {\cal L}|_{C=0} = {1\over 2 } R - 3\alpha {\partial T \partial \bar T\over (T+\bar T)^2}- {1\over 3} {F(T) F(\bar T)\over (T+\bar T)^{3 \alpha -1} } \ .
\ee
The $SL(2. \mathbb{R})$ symmetry associated with the \K\, geometry of this model can be seen as an invariance of the kinetic term (for an arbitrary $\alpha$) under the transformations

\be
T'= {aT+b\over cT+d}\ , \qquad ad-cb=1 \ .
\ee

When the imaginary part of the $T$-field is stabilized, the action becomes at $T=\bar T$
\be
e^{-1} {\cal L}|_{C=0, T=\bar T= t} = {1\over 2 } R - {3\over 4} \alpha \Big ({\partial  T\over  T} \Big )^2- {1\over 12} \tilde f^2 (T)   \,.
\ee
In canonical variables $
T= e^{\sqrt{2\over 3\alpha}\vp}$ it reads
\be
e^{-1} {\cal L}|_{C=0, T=\bar T=  e^{\sqrt {2/3\alpha} \vp}} = {1\over 2 } R - {1\over 2}  (\partial \vp)^2 - V(\vp) \ ,
\ee
where (ignoring a numerical rescaling of the potential)
\be
V(\vp)=  \tilde f^2 \Big (e^{\sqrt {2\over 3\alpha}\,  \vp}\Big )= f^2\Big(\tanh {\varphi\over\sqrt{6\alpha}}\Big).
\ee
Thus \rf{sugra} constitutes a rather elegant form of  the $\alpha$-attractor class of  models   \cite{Kallosh:2013yoa} as derived in  \cite{Cecotti:2014ipa}. Here they are defined in a complex moduli space on a half-plane $T+T>0$ instead of the disc $Z^2<1$  used in \cite{Kallosh:2013yoa}.  

It is interesting that in the half-plane variables we can look at the special case of our general models \rf{sugra} with 
$
 W=  C T^{3 (\alpha - 1)/2} (T- 1)
$.
The bosonic part is
$
V= V_{0} \big(1- e^{-\sqrt {2\over 3\alpha} \varphi}\big)^2  
$.
For  $\alpha = 1$ this model is  a Cecotti model \cite{Cecotti:1987sa} which in the bosonic case coincides 
with the potential  of  \cite{Starobinsky:1980te}. In our generic $\alpha$ supergravity case \rf{sugra} this model interpolates between  \cite{Starobinsky:1980te} at $\alpha=1$ and $\vp^2$ chaotic model \cite{Linde:1983gd} as shown in Fig.~\ref{fig:alpha1}. The case with the simpler Ansatz $W = C (T-1)$ was studied in \cite{Scalisi}; it coincides with the above for the special values $\alpha=1/3$ or $1$. 

\subsection{The Superconformal Action}

The superconformal action for the $\alpha$-attractor models using the superfield form was derived in  \cite{Cecotti:2014ipa} 
\begin{align}\label{alpha}
-\bigg[ \bar X^0 X^0 \left(T + \bar T   -  C \bar C  
 + 3 g \frac{( C \bar C)^2}{T + \bar T}\right)^{\!\!\alpha}\, \bigg]_D+\Big(\big[C F(T)(X^0)^3\big]_F+h.c.\Big).
\end{align}
When extra local symmetries of this action are fixed, one finds a supergravity model \rf{sugra}.

The scalar curvature $R_k$ of the \K\, manifold for the \rf{sugra} models is at $C=0$ is 
\be\label{curv1}
R_k\Big|_{C=0}= - {2(1-2g)\over \alpha}.
\ee
However, since the field $C$ is massive, the relevant curvature is not the scalar curvature of the K\"ahler manifold $R_k$ but  the holomorphic sectional curvature 
\begin{equation}\label{rrrffds}
 \left.\frac{R_{T\bar TT\bar T}}{(G_{T\bar T})^2}\right|_{C=0}= -\frac{2}{3\alpha},
\end{equation}   
which agrees with the scalar curvature computed using the induced metric on the half--plane $C=0$, 
\begin{equation}
G_{T\bar T}^\text{ind}=-3\alpha\,\partial_T\partial_{\bar T}\ln(T+\bar T),
\end{equation}
as explained in \cite{Cecotti:2014ipa}. Either way, at large $\alpha$ both type of curvatures tend to zero.

\subsection{Interpretation of $\alpha$}

The bosonic model \rf{action1} has a parameter $\alpha$ which has  the interpretation of regulating the difference between $\phi$ and $\varphi$. This distinction is very similar to that between velocity and rapidity in special relativity, as discussed in the context of conformal symmetry gauge-fixing in  \cite{Kallosh:2013hoa}.  The region of  very small $\alpha$  corresponds to a large rapidity, when $\tanh {\varphi\over\sqrt{6\alpha}}$ is significantly smaller than ${\varphi\over\sqrt{6\alpha}}$. At small $\alpha$, the  difference between rapidity and velocity tends to disappear:
\be
\alpha \rightarrow 0\ , \qquad \tanh {\varphi\over\sqrt{6\alpha}}\, \ll \, {\varphi\over\sqrt{6\alpha}} \ , \qquad { \rm ``ultra\,  relativistic " \,  \, limit.}
\ee
In contrast, at large $\alpha$ one has:
\be
\alpha \rightarrow \infty\ , \qquad \tanh {\varphi\over\sqrt{6\alpha}}\, \approx  \, {\varphi\over\sqrt{6\alpha}} \ , \qquad { \rm ``non\,  relativistic " \, \, limit.}
\ee
The superconformal $\alpha$-attractor models suggest an interesting interpretation of the interpolating parameter $\alpha$ in this class of models as related to a curvature of the \K\, moduli space. It is the curvature of the $SU(1,1)\over U(1)$ symmetric space with constant curvature 
\be
R_k= -{2\over 3\alpha} \ ,
\ee
 or its generalizations. This means that in the supersymmetric case the geometry of the \K\, space  has
 \be
\alpha \rightarrow 0 \ , \qquad  \qquad { \rm high\,  \, R_k \, \, curvature \,  \, limit},
\ee
or 
\be
\alpha \rightarrow \infty \ , \qquad  \qquad { \rm low \,  \, R_k \, \, curvature \, \, limit}.
\ee
The discovery of gravitational waves by BICEP2, if confirmed,  suggests that our $\alpha$-attractor models are compatible with the data at low curvature limit of the \K\, geometry.

\section{Conclusions}

While we are waiting for further guidance from Planck and BICEP2 on the preferred values of the cosmological parameters, it is interesting to concentrate on the problems which emerge with the interpretation of observational data from both of these two sources.  The models at the sweet spot of the Planck data include the Starobinsky model, the Higgs inflation  model, and a broad class of cosmological attractor models. Inflation in most of these models (except $\alpha$-attractors for $\alpha \ll 1$)  occurs at $\phi \gg 1$, so most of them belong to the class of large field chaotic inflation models, despite the fact that the amplitude of the tensor modes in these models can be extremely small. The results of BICEP2, indicating a possible contribution of the gravitational waves with $r\sim 10^{{-1}}$, also favor various large field models.

Some authors argued that large field inflation does not fit into modern physics because of a cutoff at $\phi \sim \Lambda <1$. This argument does not necessarily apply to the simplest quantum field theory models. The cutoff, if it exists, should be associated with observable quantities such as masses of particles $\sim g \phi$, rather than with the value of the inflaton field  \cite{Linde:1983gd}. However, a cutoff in moduli space may appear in theories such as supergravity and string theory, where the scalar field may have an independent geometric meaning. For this and other reasons, some authors  interpreted the recent BICEP2 data as an experimental evidence against supersymmetry and supergravity, see e.g. \cite{Lyth:2014yya}.

In this respect, it is important to notice that supergravity in its original superconformal formulation does not have any fixed energy scale associated with it. The Planck scale appears only after gauge fixing where the conformal compensator acquires a non-zero vev.  This introduces the mass scale to the theory. 

However, in some cases where this mass scale determines the boundary of the moduli space, this boundary runs to infinity upon transformation from the original variables used in the superconformal theory to canonical scalar fields in the Einstein frame. In this case, the existence of the cutoff in the original theory is not dangerous. On the contrary, it plays an important positive role: The observational predictions of such theories rapidly converge to universal values which practically do not depend on many details of the original models. In this paper, we discussed a particular class of such models, called $\alpha$-attractors. In the limit of small $\alpha$, predictions of these theories are very similar to predictions of the Starobinsky model and Higgs inflation, which lead to very small value of $r \sim 4\times 10^{{-3}}$. In the limit $\alpha \to 0$, 
the value of $r$ for $\alpha$-attractors  can go all the way down to $r= 0$, which serves as an attractor point for these theories.

In this paper we have shown that this class of model has another attractor point: In the limit $\alpha \to \infty$, the predictions of such models converge to predictions of the simplest chaotic inflation model with a quadratic potential, see Fig.~\ref{fig:alpha}.

Thus, we presented a double-attractor system, the first attractor point most favored by Planck 2013 and the second one most favored by BICEP2.  In our models a  parameter which interpolates between these two attractor points  is $\alpha$. 
Once we know the values of $n_s$ and $r$, we will be able to measure $\alpha$ in this class of theories. Therefore it is important to understand the meaning of this parameter.
 
Our bosonic models \rf{action1}  are based on the underlying superconformal/supergravity models, where the parameter $\alpha$ defines the curvature of the \K\, manifold during inflation. In our class of models, it is a symmetric  space ${SU(1,1)\over U(1)}$, where $T, \bar T$ are coordinates of the \K\, manifold with the potential $K=-3\alpha \ln(T+\bar T)$ and the single component of the \K\, metric is $g_{T\bar T} \equiv g={ 3\alpha \over (T+\bar T)^{2}}$ so that
\be
ds^2=   { 3\alpha \over (T+\bar T)^{2}} dT d\bar T \ .
\ee
This metric defines a symmetric space of a constant $T$-independent curvature 
 \be
R_{k}= g^{-3} (  \partial_T g \partial_{\bar T} g -g \partial_T\partial_{\bar T} g ) = -{2\over 3\alpha}\, .
 \ee
The situation here is very similar to what happens in inflationary theory, where the exponential growth of the scale factor $a(t)$ makes the universe flat. Here, the flatness of \K\ geometry appears in the large $\alpha$ limit.

In the context of our class of supergravity models defined in \rf{sugra},  for the choice of $F(T)$ made in this paper, we will be able to say the following:  the measurements of the curvature of the \K\, manifold $R_k= -{2\over 3\alpha}$  during inflation was performed, and the curvature must be very small to fit the BICEP2, or large to fit the models favored by Planck 2013.

From the point of view of inflationary model building in supergravity, the main class of models that was available for a long time was the class of models where the \K\ potential did not depend on one of the directions in the moduli space, which was associated with the inflaton field \cite{Kawasaki:2000yn}-\cite{Kallosh:2014xwa}. As we see now, there are two more classes of large field inflationary models in supergravity which are based on a different principle. One of them works best in the limit where the original formulation of the theory has a small cutoff, which leads to model-independent predictions with tiny values of $r$. Another one leads to equally universal predictions in the limit of the large value of the cutoff, $\Lambda \to \infty$, which in our case corresponds to the limit $\alpha \to \infty$. To discover the existence of such models, we used the first of the two approaches outlined in the Introduction: We considered the simplest models with lots of symmetries and tried to preserve their good consequences as long as possible. One might say that the existence of double attractors is not unexpected: every good story needs a villain, every attractor needs a repeller - experiment will tell where inflation has ended up and hence which of the two ends is the real attractor. We might even find that the truth is somewhere in between.

\section*{Acknowledgements}

We acknowledge stimulating discussions with Marco Scalisi, Eva Silverstein, Bert Vercnocke, Alexander Westphal and  Timm Wrase. RK and AL are supported by the SITP and by the NSF Grant PHY-1316699 and RK is also supported by the Templeton foundation grant `Quantum Gravity Frontiers'.

\end{document}